\newcommand{\detadphi}    {\ensuremath{\Delta\eta\Delta\varphi}}
\newcommand{\pp}    {\ensuremath{pp}}

 \documentclass[final,1p,times]{elsarticle}

\usepackage{amssymb}

\usepackage{lineno}

\biboptions{sort&compress}

\journal{Nuclear Physics A}

\begin{document}
\begin{frontmatter}
\title{Angular correlations of identified charged particles measured in \pp~collisions by ALICE at the LHC}
\author{$\L$ukasz Kamil Graczykowski, Ma$\l$gorzata Anna Janik (for the ALICE Collaboration)}
\address{Faculty of Physics, Warsaw University of Technology, Koszykowa 75, 00-662 Warsaw, Poland}

\begin{abstract}
We report on studies of untriggered two-particle angular correlations of identified particles (pions, kaons and protons) measured in proton-proton collisions at center-of-mass energy $\sqrt{s}=7$~TeV recorded by ALICE at the LHC. These type of studies are sensitive to a wide range of correlations which arise from different physics mechanisms, each of them having a unique structure in \detadphi~space. The correlations of particles with different quark content and flavor are sensitive to various conservation laws. The study of these correlations is the main goal of this analysis. The results confirm that these laws strongly influence the shape of the correlation functions for different particle types and must be taken into account while analysing the data. Moreover, we verify their implementation using two Monte Carlo event generators and we found that the analyzed models do not reproduce the measured correlations for protons.




\end{abstract}




\end{frontmatter}


\section{Introduction}
\label{sec:intro}
Proton-proton collisions at center-of-mass energy ${\sqrt{s}=7}$~TeV have been recorded by A~Large Ion Collider Experiment (ALICE) at the Large Hadron Collider (LHC) in 2010. They provide a unique opportunity to study Quantum Chromodynamics (QCD) at the new energy scales. The two-particle angular correlations have proven to be a robust tool which allows to explore underlying physics phenomena of particle production in collisions of both protons and heavy ions. 

\begin{figure}[!ht]
\centering
\includegraphics[width=12cm]{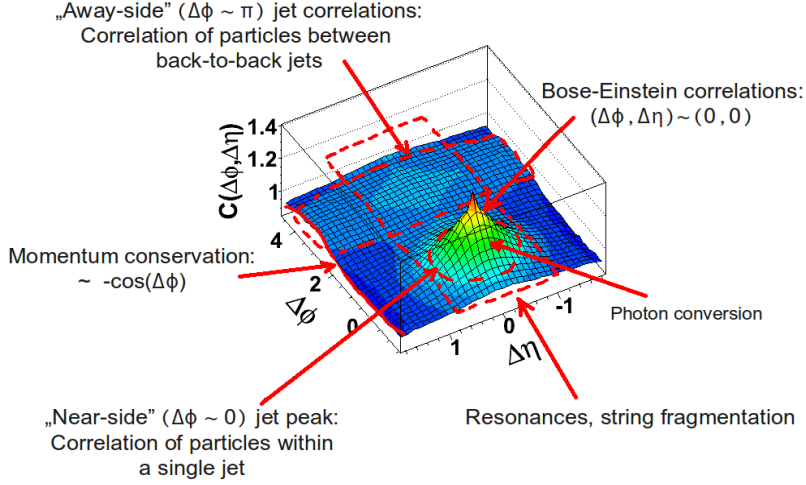}
\caption{Contributions from different correlation sources to the \detadphi~correlation function for like-sign particle pairs ($\sqrt{s}=7$~TeV \pp~collision data).}
\label{fig:ALICE_all_correlations}
\end{figure}

The angular correlations open up the possibility to study different physics mechanisms at the same time, i.e: minijets, elliptic flow, Bose-Einstein correlations, resonance decays etc. Each of these effects is a manifestation of a distinct correlation source and produces a unique distribution in \detadphi~space (where $\Delta\eta$ is the pseudorapidity difference and $\Delta\varphi$ is the azimuthal angle difference of two particles). The obtained result is a combination of all of them. The influence of all of these effects on \detadphi~correlation function can be seen in Fig.~\ref{fig:ALICE_all_correlations}. The study of untriggered non-identified particles in \pp~collisions has been performed in order to decompose the experimental correlation function and quantify the contribution of each of the correlation sources \cite{Janik:2012ya,Janik:2014cua,GraczykowskiPresentationIS2013}. The results of the analysis of triggered angular correlations in p--Pb collisions at $\sqrt{s_{\mathrm{NN}}}=5.02$~TeV and Pb--Pb collisions at $\sqrt{s_{\mathrm{NN}}}=2.76$~TeV have been reported by the ALICE Collaboration in \cite{Aamodt:2011by,Abelev:2012ola}. Similar studies were also performed for \pp~and heavy-ion systems by other experiments at LHC and RHIC in \cite{ATLAS:2012ap,Aad:2012gla,Khachatryan:2010gv,CMS:2012qk,Porter:2005rc,Alver:2007wy}.

The main goal of the analysis of untriggered angular correlations of identified particles (pions, kaons and protons -- each of them having a different quark content and flavor) is to complement the results from the studies of non-identified particles. It is especially important because in addition to the correlation sources mentioned above, we expect also the conservation laws to play a relatively significant role in the determination of the shape of the \detadphi~correlation function. Moreover, the comparison of measured correlation functions with the ones obtained from Monte Carlo event generators can be useful for the correct implementation of conservation laws in the models.



\section{Data selection}
\noindent
\emph{Event and track selection} This study was performed over minimum bias 
\pp~events at a center-of-mass energy $\sqrt{s}=7$~TeV recorded by ALICE~\cite{Aamodt:2008zz}. Two subsystems were used for particle trajectory reconstruction: the Inner Tracking System (ITS) and the Time Projection Chamber (TPC). Additionally, the VZERO detectors were used for triggering and as the multiplicity estimators and Time-Of-Flight (TOF) in the procedure of particle identification. The minimum bias trigger required a signal in either of the two VZERO counters or one of the two inner layers of the Silicon Pixel Detector (SPD).

The primary tracks were chosen within the pseudorapidity range $|\eta|<1.0$. Due to the limited detector efficiency we selected protons with the transverse momentum $p_{\mathrm{T}}>0.5$~GeV/$c$, kaons with $p_{\mathrm{T}}>0.3$~GeV/$c$ and pions with $p_{\mathrm{T}}>0.1$~GeV/$c$. Furthermore, a $p_{\mathrm{T}}$-dependent cut on the Distance of Closest Approach (DCA) was applied. Particles located within $\left( 0.018+0.035 p_{\mathrm{T}} ^{-1.01} \right)$ cm in the transverse plane and $2.0$~cm in the beam direction with respect to the primary vertex were accepted. This cut was optimized in order to reject the secondary particles. The minimum number of TPC clusters corresponding to a given track was set to 70 (maximum number of clusters in a track is 159, the number of TPC padrows). The maximum value of $\chi^2$ per TPC cluster was $4.0$ (2 degrees of freedom per cluster).
\\

\noindent
\emph{Particle identification} The particle identification (PID) of pions, kaons and protons was performed track-by-track using information from the TPC and TOF detectors. It was based on the number of sigmas ($\mathrm{n}\sigma$) method, where sigma ($\sigma$) is the standard deviation from the Bethe-Bloch parameterization of $\mathrm{d}E/\mathrm{d}x$ energy loss signal of analyzed particle in TPC or the time-of-flight deviation from the analyzed particle expected arrival time in TOF. In order to keep high purity of the sample and maximum efficiency, the upper limit of $\mathrm{n}\sigma$ for the TPC signal was set to 3 for protons, varied from 1 to 3 for kaons and from 3 to 5 for pions; depending on the momentum of the particle. In the case of TOF the upper limit of $\mathrm{n}\sigma$ was set to 3 for protons and kaons, and varied from 2 to 3 for pions. Particles were efficiently identified (purities above 90\%) up to $p_{\mathrm{T}}$ of 2.1~GeV/$c$ for pions, up to 1.43~GeV/$c$ for kaons and up to 2.6~GeV/$c$ for protons.  
\\

\noindent
\emph{Pair selection criteria} In the case of pair selection criteria, a cut (called \emph{share fraction}) preventing the effects of merging (two tracks reconstructed as one) and splitting (one track reconstructed as two) was used. It is obtained as a ratio of shared clusters to all clusters of both tracks. All pairs sharing more than $5 \%$ clusters were rejected. Moreover, electron-positron pairs from $\gamma$ conversions were removed by applying a cut which rejects pairs constructed of one positive and one negative particle having small polar angle $\Delta\theta$ difference and an invariant mass close to $0$ (the photon rest mass). 
\\

\noindent
\emph{Monte Carlo models} The analysis of collision data was complemented with the Monte Carlo (MC) model studies. Two MC event generators were used -- PYTHIA 6.4 tune Perugia-0 \cite{Sjostrand:2006za,Skands:2010ak} and PHOJET version 1.12~\cite{Engel:1995sb}. The MC generated events were processed through the reconstruction chain of the ALICE framework in order to simulate the detector response. Such reconstructed MC data were analyzed by applying the same set of particle and pair selection criteria as for the collision data. The study showed that the influence of the detector efficiency does not alter the physics message. Moreover, the correlation functions obtained from both models are very similar. Therefore, only the results from PYTHIA 6.4 Perugia-0 generated events will be shown later.

\section{Construction of correlation function}
The measured \detadphi~correlation function includes physical correlations as well as undesired effects from non-uniform detector acceptance. In order to correct for those imperfections the correlation function is constructed as follows:
\begin{equation}
\label{eq:CorrelationFuntion}
C(\Delta\eta,\Delta\varphi)=\frac{N_{pairs}^{mixed}}{N_{pairs}^{signal}} \frac{S(\Delta\eta,\Delta\varphi)}{B(\Delta\eta,\Delta\varphi)},
\end{equation}
where $\Delta\eta=\eta_1 - \eta_2$ is the difference in
pseudorapidity, $\Delta\varphi=\varphi_1 - \varphi_2$ is the difference in
azimuthal angle, $S(\Delta\eta,\Delta\varphi)$ is called the signal distribution and $B(\Delta\eta,\Delta\varphi)$ is called the background distribution containing only the detector effects. The signal distribution is constructed from particle pairs coming from the same event:
\begin{equation}
\label{eq:Nsignal}
S(\Delta\eta,\Delta\varphi)=\frac{\mathrm{d}^2N_{pairs}^{signal}}{\mathrm{d}\Delta\eta \mathrm{d}\Delta\varphi},
\end{equation}
where $N_{pairs}^{signal}$ is the number of pairs of protons, kaons or pions. 
The background distribution is constructed using the procedure of event mixing, where each particle in the pair comes from a different event and can be expressed as:
\begin{equation}
\label{eq:Nmixed}
B(\Delta\eta,\Delta\varphi)=\frac{\mathrm{d}^2N_{pairs}^{mixed}}{\mathrm{d}\Delta\eta \mathrm{d}\Delta\varphi},
\end{equation}
where $N_{pairs}^{mixed}$ is the number of pairs of protons, kaons or pions in the background distribution $B$. In order to improve the background estimation, each event is mixed with ten different events similar in terms of multiplicity and primary vertex location.

 We note that the correlation functions shown below are not corrected for tracking efficiency as well as purity and contamination.

\section{Results}
\label{sec:results}

\begin{figure}[!h]
  \centering
  \includegraphics[width=1.0\textwidth]{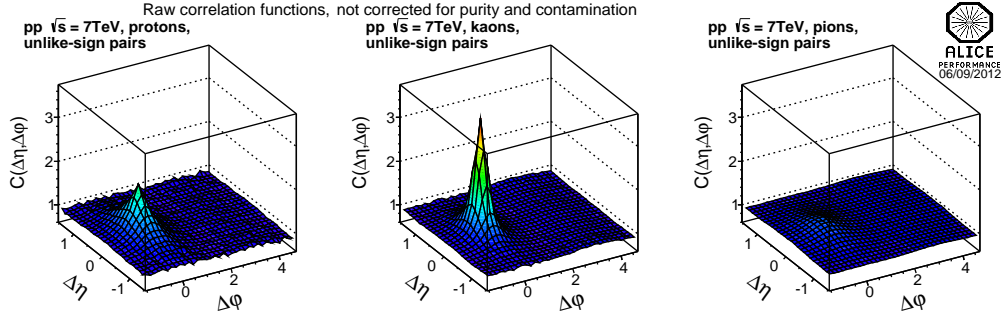}
  \caption{Correlation functions for unlike-sign pairs of protons (left), kaons (middle) and pions (right) for $\sqrt{s}=7$~TeV \pp~data.}
  \label{Fig:corrMergedMixed}
\end{figure}

\begin{figure}[!h]
  \centering
  \includegraphics[width=1.0\textwidth]{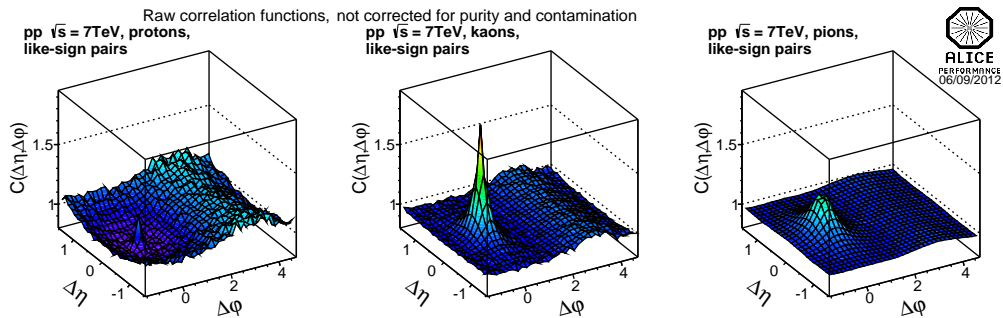}
  \caption{Correlation functions for like-sign pairs of protons (left), kaons (middle) and pions (right) for $\sqrt{s}=7$~TeV \pp~data.}
  \label{Fig:corrMergedSame}
\end{figure}

The measured correlation functions for all three analyzed particle types (pions, kaons, protons) are shown in Fig.~\ref{Fig:corrMergedMixed} for unlike-sign pairs and in Fig.~\ref{Fig:corrMergedSame} for like-sign pairs. The magnitude of the near-side peak with a center at $(\Delta\eta,\Delta\varphi) = (0,0)$ is different for different particle species. Both in the case of unlike-sign and like-sign pairs the correlation is the strongest (the near-side peak most distinct) for kaon pairs and significantly lower for proton and pion pairs. However, a surprising behavior is present in the case of protons -- for p$\mathrm{\bar{p}}$ the near-side peak is stronger than for  $\mathrm{\pi^{+}}\mathrm{\pi^{-}}$, but for like-sign pairs we do not observe a near-side peak but a wide dip with a minimum at $(\Delta\eta,\Delta\varphi) = (0,0)$. Additionally to the wide dip, a narrow correlation peak, arising from two-proton strong interaction (observed also in pp femtoscopic correlations~\cite{Salzwedel:2013aza,Szymanski:2012qu}), is located at $(\Delta\eta,\Delta\varphi) = (0,0)$. The magnitude of the away-side correlation is small in comparison to the near-side peak in the case of unlike-sign pairs for all particle types. However, it is more prominent for the like-sign pairs (stronger for protons and kaons, the weakest for pions).

We believe the following reasoning which explains these results. The production of different types of particles is governed by various conservation laws, summarized in Tab.~\ref{tab:conservLaws}. In the case of pions (the lightest hadrons with a rest mass~$0.139\ \rm{GeV}/${\it c}) only the momentum and electric charge must be conserved. In the case of charged kaons (with a rest mass~$0.494\ \rm{GeV}/${\it c}), which contain a strange quark, strangeness must be conserved. In the case of protons (the heaviest of analyzed particles with a rest mass~$0.938\ \rm{GeV}/${\it c}), the baryon number must be conserved. Since every single collision is an independent system, it must conserve all of these quantities. This is the definition of global conservation laws. Moreover, all the quantum numbers must be conserved in each parton fragmentation~\cite{Aihara:1986fy,Althoff:1984ut,Buskulic:1994ny,Abreu:1997mp,Acton:1993ux,Muller:1999yg}. We test at which scale this conservation takes place. If one assumes that the conservation laws are obeyed at a scale smaller than the whole event then they must influence the shape of the correlation function. The study of different particle species, where different conservation laws are present, should allow to determine that.


\begin{table}[!h]
\centering
\caption{The conservation laws and quantum numbers that are present for different particle types.}
\begin{tabular}{ccccc}
\hline
particles	&momentum	&charge	&strangeness	&baryon number\\\hline
pions	& \checkmark	& \checkmark	&	        &\\
kaons	& \checkmark	& \checkmark	& \checkmark	&\\
protons	& \checkmark	& \checkmark	&              	& \checkmark\\\hline
\end{tabular}
\label{tab:conservLaws}
\end{table}

We first focus on the unlike-sign particle pairs which in this analysis were constructed in all cases with a particle and its antiparticle. It is always energetically most favorable to compensate particles with their antiparticles. The strength of the correlation for such a pair depends on the energetic price of an alternative solution (if the alternative solution is ``cheap'' then the correlation will be weaker than for the ``expensive'' alternative solutions, where the correlation will be stronger):
\\(1) For pions the alternative solution is any opposite-charge particle. Since there are a lot of them in the event the cost of such solution is not very high and therefore the magnitude of the near-side peak is the lowest.
\\(2) For protons another antibaryon (charged, or neutral plus additional charged particle) would have to be produced. This solution is less probable (the energetic price is higher) than the alternative solution in $\pi^{+}\pi^{-}$ case which results in a stronger near-side correlation.
\\(3) For kaons, which carry a strange quark, strangeness must be conserved; so, the alternative solution would be at least a lambda together with another baryon. The energetic price of such a solution is very high and that is why the near-side peak for kaons is the strongest.

On the other hand, for identical particle pairs the effect is different. In that case the masses of particles play a significant role. In addition, the femtoscopic effects (Bose-Einstein for identical bosons or Fermi-Dirac quantum statistics for identical fermions) which increase or decrease the near-side correlation have to be also taken into account. We will consider the pp case where a dip is observed. If the conservation laws were obeyed only globally, then the momentum and baryon number would be compensated by two antiprotons going in the opposite direction. Such a solution has moderate energetic price which produces a relatively low correlation. If these quantities in addition should be conserved locally then while producing two protons, two very heavy particles, in the same direction we would have to produce another two antibaryons to compensate the baryon number. Moreover, in order to compensate the momentum, additional particles would have to be produced going in the opposite direction. Such a solution is an extremely expensive scenario, potentially producing a dip.

\begin{figure}[h]
  \centering
  \includegraphics[width=1.0\textwidth]{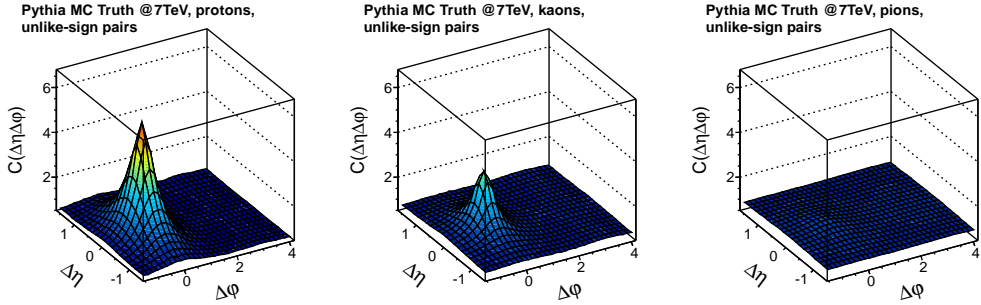}
  \caption{Correlation functions for PYTHIA 6.4 Perugia-0 $\sqrt{s}=7$~TeV \pp~data for unlike-sign pairs of protons (left), kaons (middle) and pions (right). }
  \label{Fig:PythiaMixed}
\end{figure}

\begin{figure}[h]
  \centering
  \includegraphics[width=1.0\textwidth]{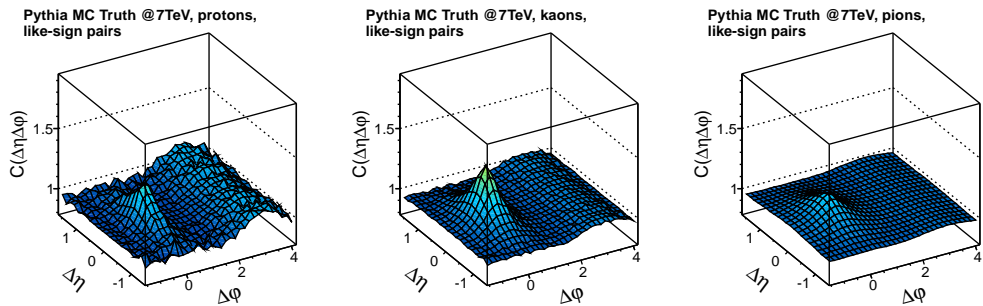}
  \caption{Correlation functions for PYTHIA 6.4 Perugia-0 $\sqrt{s}=7$~TeV \pp~data for like-sign pairs of protons (left), kaons (middle) and pions (right).}
  \label{Fig:PythiaSame}
\end{figure}

The experimental data were compared to Monte Carlo models -- in Fig.~\ref{Fig:PythiaMixed} the correlation functions for unlike-sign particles and in Fig.~\ref{Fig:PythiaSame} for like-sign pairs from PYTHIA 6.4 Perugia-0 are shown. There are striking differences between these results and the ones from collision data in Fig.~\ref{Fig:corrMergedMixed} and Fig.~\ref{Fig:corrMergedSame}. First of all, for the unlike-sign pairs the strongest correlation is observed for protons, not kaons -- the near-side peak for p$\bar\mathrm{p}$ is practically twice as strong as the strength for $\mathrm{K^{+}K^{-}}$. Regarding the away-side correlation, it is relatively similar to the collision data. In the case of like-sign pairs again the protons are the most different. First of all we do not observe any dip but the near-side peak is present. Moreover, the significant away-side correlation is pronounced with the magnitude comparable to the strength of the near-side correlation.

In summary, the strength and the character of identical proton correlations suggest that the conservation laws play a major role at a scale significantly smaller than the whole event. On the other hand, this seems not to be precisely modeled in Monte Carlo generators tested (PYTHIA, PHOJET). Similarly as in the collision data, the near-side peak is most promiment for kaons in PYTHIA. The correlation functions for pions show the most similar behavior as for ALICE data both for like-sign and unlike-sign pairs.





\section{Conclusions}
\label{sec:conclusions}
The analysis of angular correlations of identified particles in \detadphi~space in \pp~collisions at $\sqrt{s}=7$~TeV with ALICE was performed. The studies were done separately for different charge combinations of particle pairs (for like-sign and unlike-sign pairs) and for three particle types (pions, kaons, protons). The analysis was also complemented by the Monte Carlo model studies using PYTHIA 6.4 Perugia-0 and PHOJET event generators. The results show that the shape of the correlation function is different for each particle type; the strongest correlation being observed for kaons. A significant dip instead of a near-side peak was observed for the like-sign proton pairs. The results from Monte Carlo model studies show surprisingly different features in the correlation functions for protons, both for like-sign pairs (where we do not observe a dip in MC data) and unlike-sign pairs (where the near side correlation is the strongest for protons).  

We propose the possible mechanism which could explain the observed shapes of \detadphi~correlation functions and the discrepancies between experimental data and Monte Carlo models. It is based on the fact that the local conservation laws (obeyed for each parton fragmentation), such as charge, strangeness or baryon number, play a major role in the determination of the correlation function shape. We stress that this explanation should be verified by dedicated model studies and these results could potentially constrain the Monte Carlo event generators on how the conservation laws are implemented.

\section*{Acknowledgements}
\label{sec:acknowledgements}
This work has been financed by the Polish National Science Centre under decisions no. DEC-2011/01/B/ST2/03483, DEC-2012/05/N/ST2/02757, and by the European Union in the framework of European Social Fund.
%



\bibliographystyle{elsarticle-num}
\bibliography{bibliography}







\end{document}